# Anomalous Hall hysteresis in $Tm_3Fe_5O_{12}$/Pt with strain-induced perpendicular magnetic anisotropy

Chi Tang[1], Pathikumar Sellappan[2], Yawen Liu[1], Yadong Xu[1], Javier E. Garay[2], and Jing Shi[1]

[1]Department of Physics and Astronomy, University of California, Riverside, CA 92521, USA

[2]Department of Mechanical Engineering and Materials Science & Engineering Program, University of California, Riverside, CA 92521, USA

We demonstrate robust interface strain-induced perpendicular magnetic anisotropy in atomically flat ferrimagnetic insulator $Tm_3Fe_5O_{12}$ films grown with pulsed laser deposition on substituted-$Gd_3Ga_5O_{12}$ substrate which maximizes the tensile strain at the interface. In bilayers consisting of Pt and TIG, we observe large squared Hall hysteresis loops over a wide range of thicknesses of Pt at room temperature. When a thin Cu layer is inserted between Pt and TIG, the Hall hysteresis magnitude decays but stays finite as the thickness of Cu increases up to 5 nm. However, if the Cu layer is placed atop Pt instead, the Hall hysteresis magnitude is consistently larger than when the Cu layer with the same thickness is inserted in between for all Cu thicknesses. These results suggest that both the proximity-induced ferromagnetism and spin current contribute to the anomalous Hall effect.

The ferrimagnetic insulator yttrium iron garnet (YIG) has continually attracted a great deal of attention in condensed matter physics, especially in the spintronics community. For example, the discovery of the spin Seebeck effect has led to a rapid development of insulator-based spin-caloritronics[1, 2]. In bilayer structures containing a normal metal (NM) layer having strong spin-orbit coupling and a YIG layer, magnetoresistance in Pt emerges and evolves with temperature[3-11]. Associated with many interesting phenomena in YIG/NM bilayers, a hotly debated issue is whether the magnetic proximity effect or the pure spin current effect plays a more important role. Both mechanisms can produce magnetoresistance, namely the anisotropic magnetoresistance (AMR), the spin Hall magnetoresistance (SMR), and the anomalous Hall-like effects. A proximity-induced ferromagnetic layer in Pt can generate the anomalous Hall effect (AHE) just as normal ferromagnetic conductors do. However, pure spin current, through the non-zero imaginary part of the spin-mixing conductance, can also give rise to an AHE-like response at the YIG/NM interface[12], but the relative importance of each mechanism has not been systematically addressed.

Experimentally, anomalous Hall-like response has been observed in YIG/NM bilayers such as YIG/Pt and YIG/Pd[4, 6, 10, 13]. In ferromagnetic conductors, the Hall response contains two parts: the ordinary Hall effect (OHE) which is linear in field, and the AHE which is proportional to the out-of-plane magnetization. Since YIG grown on gadolinium gallium garnet (GGG) has easy-plane anisotropy, the AHE signal is non-hysteretic but saturates at high fields. This has been treated as the basis of separating out the AHE contribution. However, such separation can be problematic. First, the observed Hall saturation field is an order larger than that of the YIG magnetization (~2000 Oe)[4, 6, 10]. Second, even the sign and magnitude of the OHE background of YIG/NM is far from being understood[13, 14]. If a ferrimagnetic insulator has perpendicular magnetic anisotropy (PMA), the squared Hall hysteresis loop in bilayers would eliminate the aforementioned problem and identify the Hall signal associated with the magnetization, i.e. the AHE. In this letter, we first demonstrate the synthesis of both $Tm_3Fe_5O_{12}$ (thulium iron garnet or TIG) thin films with strong PMA by controlling interface tensile strain and then bilayers of TIG/Pt. We show squared Hall hysteresis loops in TIG/Pt bilayers with a range of Pt thicknesses. Since Cu is known to have a very long spin diffusion length (~350 nm at room temperature[15]), we insert a thin Cu layer between TIG and Pt to suppress the proximity coupling while allowing spin current to propagate. We further investigate the effect of the Cu layer thickness on the Hall hysteresis.

Although previous studies[16, 17, 18, 19] on YIG films indicated the existence of finite PMA due to stain, a systematic study is still lacking. Here we leverage the lattice-mismatch induced strain in epitaxially grown films to generate a perpendicular surface anisotropy field $H_\perp$. When the surface anisotropy is sufficiently strong to overcome the shape anisotropy, a squared hysteresis loop is resulted[20]. Along the <111> orientation of cubic crystals[21], $H_\perp = \frac{-4K_1 - 9\lambda_{111}\sigma_\|}{3M_s}$, where $K_1$, $\lambda_{111}$, $\sigma_\|$, and $M_s$ stand for the 1st-order cubic anisotropy, magnetostriction constant, in-plane stress, and saturation magnetization of the film, respectively. TIG, as a member of the rare earth iron garnet family, has a large negative magnetostriction constant ($\lambda_{111}$= -5.2) [22], about twice as much as that of YIG ($\lambda_{111}$= -2.7) [22]. A large tensile strain is needed to produce a large positive $H_\perp$. The most commonly used GGG substrate has a lattice constant of 12.383 Å while the substituted-gadolinium gallium garnet (SGGG) substrate has a lattice constant of 12.585 Å, larger than TIG (a=12.324 Å) or YIG (a=12.376 Å). Thereby we choose SGGG/TIG to maximize the tensile strain at the interface.

Pulsed laser deposition (PLD) is well known for growing coherently strained pseudomorphic films[15]. We first prepare our TIG PLD targets. Highly homogeneous and dense targets are fabricated through combination of a chemical precursor approach and the current activated and pressure assisted densification method[23, 24]. The detailed information about TIG target preparation will be published elsewhere[23]. We have successfully grown atomically flat, epitaxial TIG thin films on (111)-oriented SGGG substrates. After going through standard cleaning, the substrates are annealed at ~ 200 ºC for over five hours prior to TIG deposition. Under the optimal growth condition, the as-grown TIG films exhibit strong PMA and ultra-flat surfaces, as shown in Fig 1. Fig. 1(a) displays both in-plane and out-of-plane magnetic hysteresis loops of a typical 10 nm thick TIG film measured by a vibration sample magnetometer (VSM). The out-of-plane hysteresis loop is squared while the in-plane hysteresis loop shows a hard-axis behavior. The in-plane saturation field $H_{sat}$ is ~2000 Oe. From the spontaneous magnetic moment per unit cell (~2.8 $\mu_B$ at room temperature[25]) and the volume of the unit cell, $4\pi M_s$ is found to be ~1393 G. From $H_\perp = H_{sat} + 4\pi M_s$, $H_\perp$ is calculated to be 3393 Oe. On the other hand, $H_\perp$ can be estimated from material parameters by $H_\perp = \frac{-4K_1 - 9\lambda_{111}\sigma_\|}{3M_s}$. Since the 1st-order anisotropy is known to be smaller, $H_\perp$ can be approximated by $-3\lambda_{111}\sigma_\parallel/M_s$. Based on the elastic stiffness of TIG bulk crystals and lattice-mismatch induced in-plane strain, the in-plane tensile stress[26] $\sigma_\parallel$ is estimated to be $1.4 \times 10^4$ erg/cm$^3$. Using the magnetostriction constant $\lambda_{111}$ = -5.2 for TIG bulk crystals, we obtain $H_\perp \sim$ 1969 Oe, somewhat lower but on the same order as the experimental value. The discrepancy may originate from the underestimation of the in-plane tensile stress based on the bulk crystal magnetostriction constant and the interfacial strain estimated from the lattice constants of the two bulk crystals. As a comparison, we also grow a 10 nm YIG film on GGG (111) under the same growth condition. The inset of Fig. 1(a) displays magnetic hysteresis loops of YIG on GGG, which clearly show in-plane anisotropy due to dominant shape anisotropy. Fig. 1(b) is the atomic force microscopy (AFM) image of a representative TIG film. The root-mean-square (rms) roughness over a 10 μm × 10 μm scanned area is ~ 1.4 Å. All samples with the same film thickness (10 nm) show similar rms roughness.

To further characterize the magnetic anisotropy, a ferromagnetic resonance (FMR) study is carried out using a Bruker 9.3 GHz X-band EMX EPR spectrometer. A static magnetic field $H$ is applied at an angle $\theta_H$ with respect to the film normal. Fig. 2(a) displays the FMR spectra of a TIG film with $H$ applied both parallel ($\theta_H$=90°) and perpendicular to the film plane ($\theta_H$=0°). The resonance field $H_{res}$ is 1900 Oe and 5200 Oe, and the peak-to-peak linewidth $\Delta H$ is 224 Oe and 167 Oe for field out-of-plane and in-plane, respectively. $\Delta H$ of the TIG thin film is much greater than that of YIG films or YIG crystals, but is comparable with that of TIG bulk single crystals ($\Delta H$=135 Oe[27]), which may be attributed to the intrinsic properties of the materials. Note that the out-of-plane $H_{res}$ is smaller than the in-plane $H_{res}$, which is a qualitative indicator of PMA. The detailed polar angle dependence of $H_{res}$ is summarized in Fig. 2(b) for GGG/YIG, SGGG/YIG, GGG/TIG, and SGGG/TIG with different levels of tensile strain. The effective demagnetization field $4\pi M_{eff} = 4\pi M_s - H_\perp$ is extracted by fitting the resonance equation to the angular dependence of $H_{res}$[28, 29] and then $H_\perp$ is calculated to be -232 Oe, 1078 Oe, 1460 Oe, and 3645 Oe for GGG/YIG, SGGG/YIG, GGG/TIG, and SGGG/TIG, respectively. The $H_\perp$ value (3645 Oe) is the highest for SGGG/TIG which agrees well with what is obtained from magnetometry. In GGG/YIG, negative $H_\perp$ indicates in-plane anisotropy. The increasing trend of $H_\perp$ is correlated well with the increasing interface tensile strain, manifesting strain tunable PMA in rare earth iron garnet films.

To maintain high-quality interfaces in bilayers, TIG is immediately transferred to a high-vacuum sputtering system for Pt or Cu deposition without being exposed to any resist or organic substance. Before the metal layer deposition, TIG films are lightly cleaned using Ar plasma for 3 minutes. Standard photolithography and Ar inductively coupled plasma etching are performed to pattern the metal layers into Hall-bars with the length of $L$=300 μm and the width of $W$=100 μm. Magneto-transport measurements are performed at room temperature either with an electromagnet or with a superconducting magnet in a Physical Property Measurement System. Fig. 3(a) shows the Hall hysteresis loops for two representative samples: TIG (10 nm)/Pt (1.5 nm) and TIG (10 nm)/Pt (5 nm) along with the Hall signals in two reference samples: SGGG/Pt (2 nm) and TIG/SiO$_2$ (2 nm)/Pt (2 nm). Three messages are revealed in the figure: 1. the Hall hysteresis in Pt resembles the out-of-plane magnetic hysteresis of the underlying TIG film; 2. a thin SiO$_2$ layer quenches the Hall hysteresis completely; 3. the thinner the Pt layer is, the larger is the Hall hysteresis magnitude. In the reference samples, only linear OHE is present. Similar OHE linear background is also present in TIG/Pt samples, but unlike in YIG-based bilayers, there is no ambiguity in separating AHE from the total Hall signal. In samples with different Pt thicknesses but sharing the *same* underlying TIG film, the AHE magnitude steadily decreases as the Pt layer thickness as shown in Fig. 3(b), suggesting its interfacial nature.

Two possible mechanisms can give rise to the AHE hysteresis in paramagnetic Pt. First, if the Pt interface layer is magnetized by TIG via proximity coupling, then it behaves effectively as a thin ferromagnetic metal, and consequently the hysteresis can arise from the conventional AHE mechanism[30]. As the Pt thickness increases, the AHE signal from the interface layer is diluted by the increasing paramagnetic portion. The other possible mechanism is the spin Hall-AHE (SH-AHE), a spin current effect[12, 13], which originates from spin precession around the exchange field due to the presence of the magnetic layer. Both SH-AHE and SMR share the same origin, and are theoretically connected to the imaginary and real parts of the same spin-mixing conductance respectively. We have carried out room temperature magnetoresistance measurements with a rotating field 1 T in the same set of samples as used for the Fig. 3(b) inset. The SMR data show a similar decreasing trend which is included in the supplementary materials[31]. Therefore, both mechanisms seem to be plausible to explain the Pt thickness dependence of the Hall magnitude.

In order to further distinguish the two mechanisms of the Hall hysteresis in TIG/Pt, we insert a thin layer of Cu between TIG and Pt. Since Cu has a long spin diffusion length (~350 nm), a few nm thick Cu layer should not suppress the spin current; therefore, no obvious effect is expected on AHE except for current shunting, if the spin current is responsible for the Hall hysteresis. On the other hand, a thin Cu layer should significantly affect the proximity coupling and therefore the AHE magnitude if the induced magnetic interface layer is responsible. To exclude the shunting effect, we prepare two sets of samples: TIG/Pt/Cu/SiO$_2$ (1) and TIG/Cu/Pt/SiO$_2$ (2) on the same TIG film, with exactly the same constituent layer thicknesses but the opposite stacking order for the Pt and Cu layers. The SiO$_2$ capping layer is important to prevent oxidation of the top layer, especially in Set 1 when Cu is at the top. As shown in Fig. 4(a), upon insertion of a Cu spacer layer immediately above TIG, the absolute magnitude of the Hall hysteresis is quickly suppressed. What is more important, by placing a Cu layer of the same thickness on TIG/Pt instead, the AHE hysteresis is decreased by a lesser amount. Furthermore, a 5 nm Cu layer greatly reduces the Hall hysteresis in Set 2, but the AHE signal clearly remains finite. From AFM imaging, the rms roughness associated with the 2 nm Cu grown on TIG is ~ 0.15 nm (see Supplementary[32]); therefore, we rule out the possibility of pinholes in Cu films. The finite Hall hysteresis loop in Set 2 strongly suggests that spin current plays a more important role rather than the proximity coupling when the inserted Cu is thick, since the latter is expected to be short-ranged.

We vary the Cu layer thickness from $t_{Cu}$ 1.5 to 5 nm in both TIG/Pt(2 nm)/Cu/SiO$_2$ and TIG/Cu/Pt(2 nm)/SiO$_2$ sets. We note that even with the same Cu thickness the resistance of the two samples with the opposite Cu and Pt stacking orders is different. The resistance difference is more pronounced in thin Cu samples. It may be caused by different Cu textures when it is grown on different materials, i.e. TIG or Pt. To better correlate the shunting effect as the Cu layer thickness is varied, we measure the total resistance of the samples. In Fig. 4(b), we plot the AHE magnitude as a function of the total measured resistance instead of the Cu thickness. As the Cu thickness increases, the resistance of the both sets decrease and at the same time the AHE magnitude decreases due to the shunting effect. However, the TIG/Pt (2 nm)/Cu/SiO$_2$ AHE curve stays consistently above that of TIG/Cu/Pt (2 nm)/SiO$_2$. The gap between these two AHE curves reveals the importance of the magnetic proximity coupling. As the Cu thickness approaches zero in Set 2, i.e. extrapolating the curve for Set 2 to the same resistance value as that of TIG/Pt (2 nm)/SiO$_2$, the difference between the two curves, $\Delta R_{AH}^1$, should represent the contribution from the proximity effect, as shown in Fig. 4(b). After separating out the proximity-induced AHE contribution, the remaining AHE signal, $\Delta R_{AH}^2$, is clearly from the spin current effect, i.e. SH-AHE. Below we adopt a simple model to explain the overall Cu layer thickness dependence in both sets. We assume the AHE voltage from the Pt layer to be the voltage source and the presence of the Cu layer merely shunts the current flowing in Pt; therefore the measured AHE voltage is reduced. The AHE source may contain more than one mechanism. With this simple circuit model, the measured AHE resistance scales with the total sample resistance squared, i.e. $R_{AH} = \left(\frac{R}{R_{Pt}}\right)^2 R_{AH}^0$, here $R_{AH}^0$ being the AHE resistance from the Pt layer only regardless of its physical origin, $R$ being the total resistance of the samples, and $R_{Pt}$ being the resistance of the Pt layer. Fig. 4(c) is the $R_{AH}^0$ vs. $t_{Cu}$ plot. As the Cu thickness varies, $R_{AH}^0$ is nearly constant for both sample sets. It means that the decreased AHE magnitude in thicker Cu samples can indeed be described by the current shunting effect. When the Pt layer is directly on top of TIG, the constant value is clearly larger than when the Cu layer separates the Pt layer from TIG. The former contains both proximity induced AHE and SH-AHE; therefore, we attribute the difference between the two values to the proximity induced mechanism which apparently dominates the other. However, we cannot completely rule out a possibility that the attenuation of the spin current by YIG/Cu and Cu/Pt interfaces gives rise to the dramatic reduction in the AHE magnitude.

In summary, by controlling interfacial strain, we have obtained robust PMA in TIG thin films. Squared AHE hysteresis loops are observed in TIG/Pt bilayers and analyzed in the context of both proximity coupling and spin current effects. Our experimental results indicate that the former effect likely plays a dominant role in the anomalous Hall effect.

We would like to thank B. Madon, Z. L. Jiang, and J. X. Li for many helpful discussions, and N. Amos, J. Butler and D. Yan for their technical assistance. The work was supported as part of the SHINES, an Energy Frontier Research Center funded by the U.S. Department of Energy, Office of Science, Basic Energy Sciences under Award # SC0012670.

**Figure Caption:**

Fig. 1 (a) Magnetic hysteresis loops of 10 nm TIG film grown on SGGG (111) substrate for both field in-plane (ip) and out-of-plane (op) orientations showing perpendicular magnetic anisotropy. Linear paramagnetic background is removed. Inset: Magnetic hysteresis loops of 10 nm YIG film grown on GGG (111) substrate. (b) AFM surface morphology of TIG film with rms roughness of 0.14 nm.

Fig. 2. (a) FMR spectra of SGGG/TIG (10 nm) with field in-plane and out-of-plane orientations. $\theta_H$ is the angle of magnetic field with respect to the film normal. (b) Polar angle $\theta_H$ dependence of the resonance field $H_{res}$ for GGG/YIG, SGGG/YIG, GGG/TIG, and SGGG/TIG bilayers. The fitting curves (solid) and the extracted perpendicular anisotropy fields are shown.

Fig. 3. (a) Total Hall resistivity $\rho_{xy}$ for TIG/Pt(1.5 nm), TIG/Pt (5 nm), SGGG/Pt(2 nm), and TIG/SiO$_2$(2 nm)/Pt (2 nm). (b) The anomalous Hall resistivity for different TIG (10 nm)/Pt samples with different Pt layer thicknesses: 1.5, 2, 4, 6, and 8 nm. Inset: Thickness dependence of the anomalous Hall resistivity.

Fig. 4. (a) The anomalous Hall resistance $R_{AH}$ for three representative TIG/Pt(2 nm)/Cu($t_{Cu}$)/SiO$_2$(5 nm) and TIG/Cu($t_{Cu}$)/Pt(2 nm)/SiO$_2$(5 nm) samples with $t_{Cu}$=3, 4 and 5 nm on a same 10 nm thick TIG film. (b) The anomalous Hall resistance $R_{AH}$ as a function of the longitudinal resistance R for the two sets with all Cu thicknesses, i.e. $t_{Cu}$=1.5, 2, 3, 4 and 5nm. Solid curves are the fitting based on the shunting model of AHE: $R_{AH} \sim R^2$. (c) Calculated unshunted anomalous Hall resistance $R_{AH}^0$ as a function of the Cu thickness for the two sample sets.

(a)

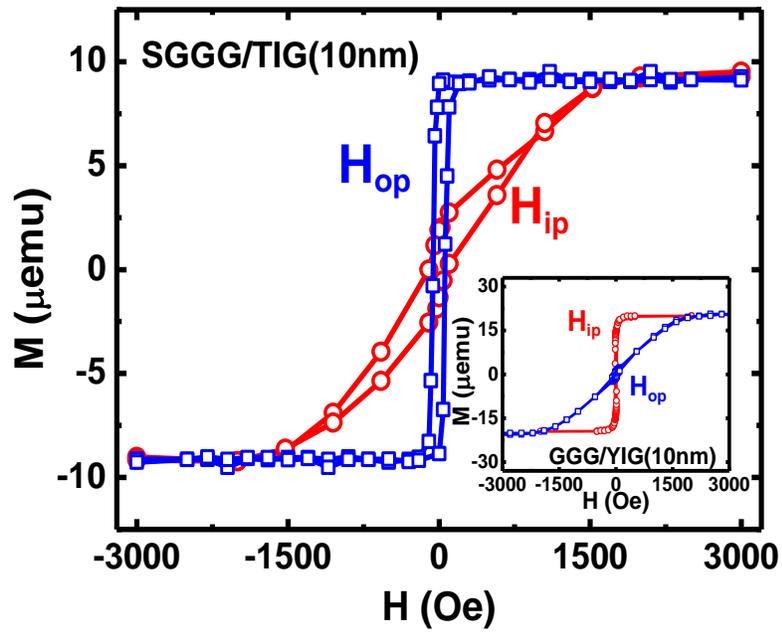

(b)

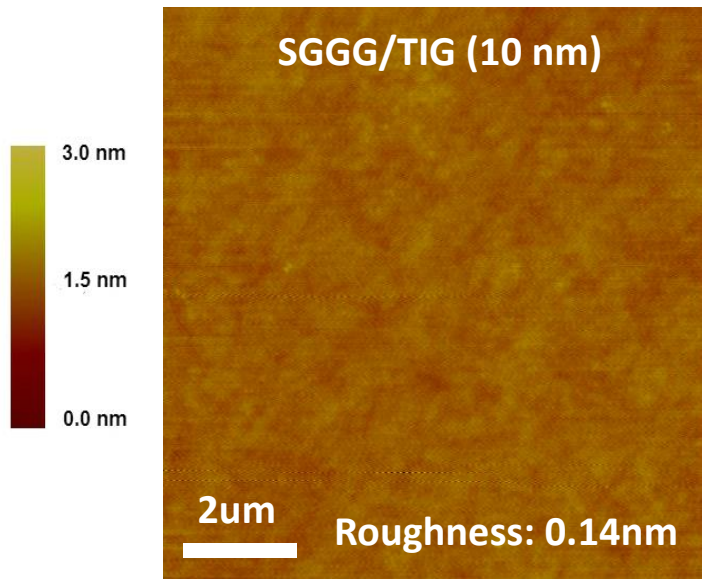

**Figure 1**

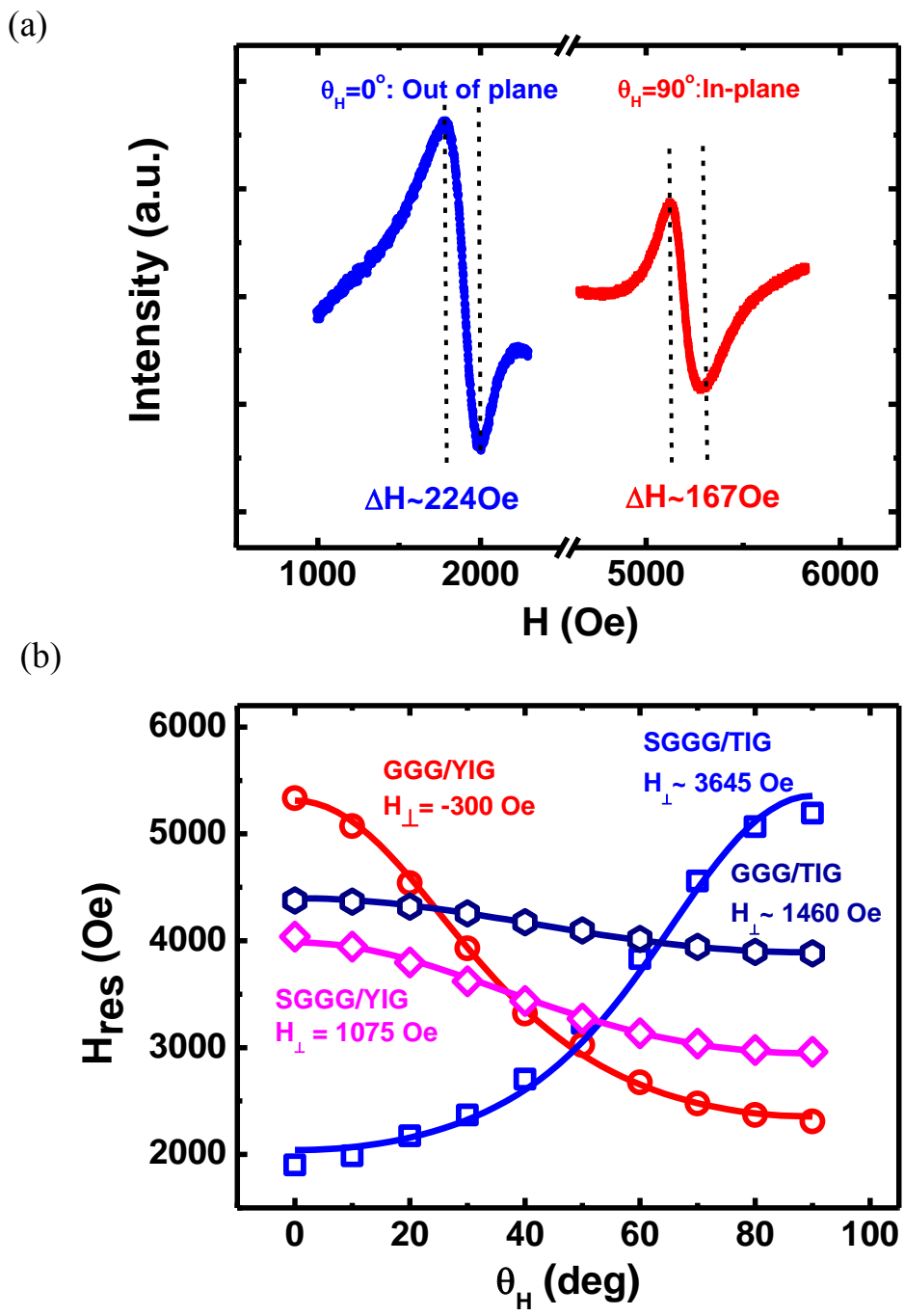

**Figure 2**

(a)

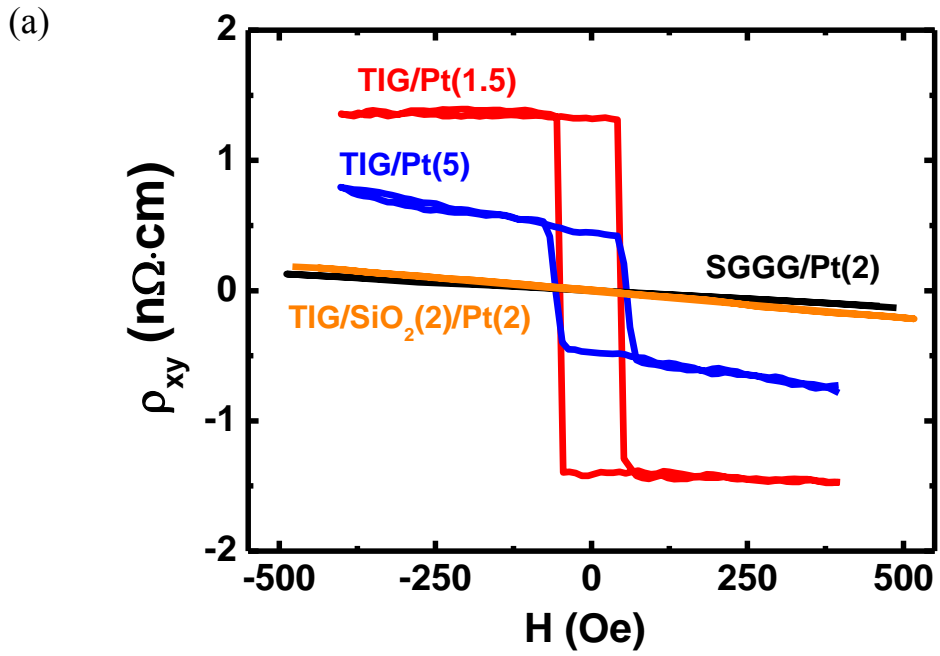

(b)

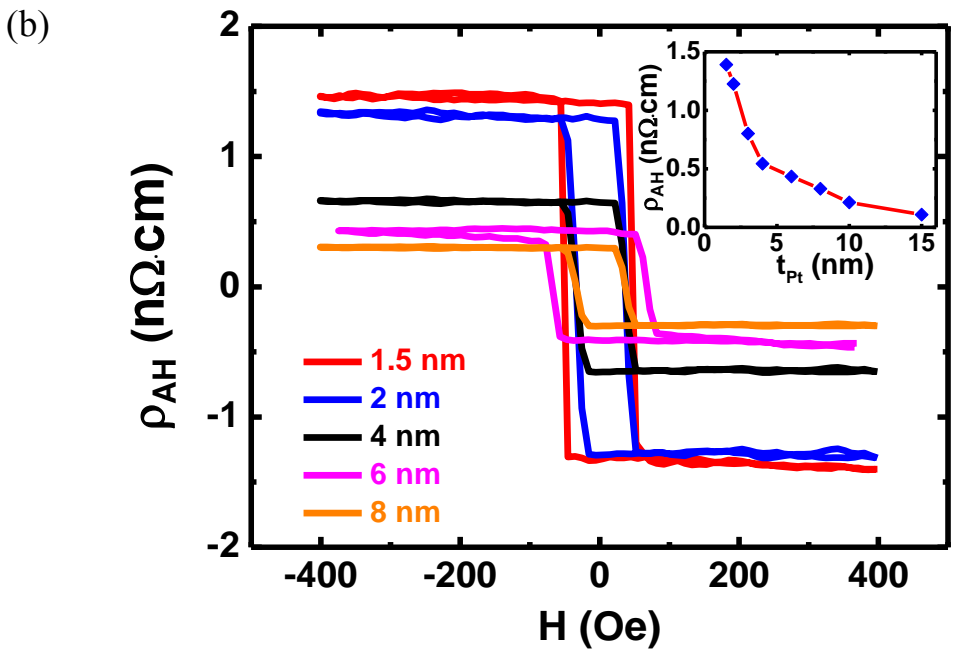

**Figure 3**

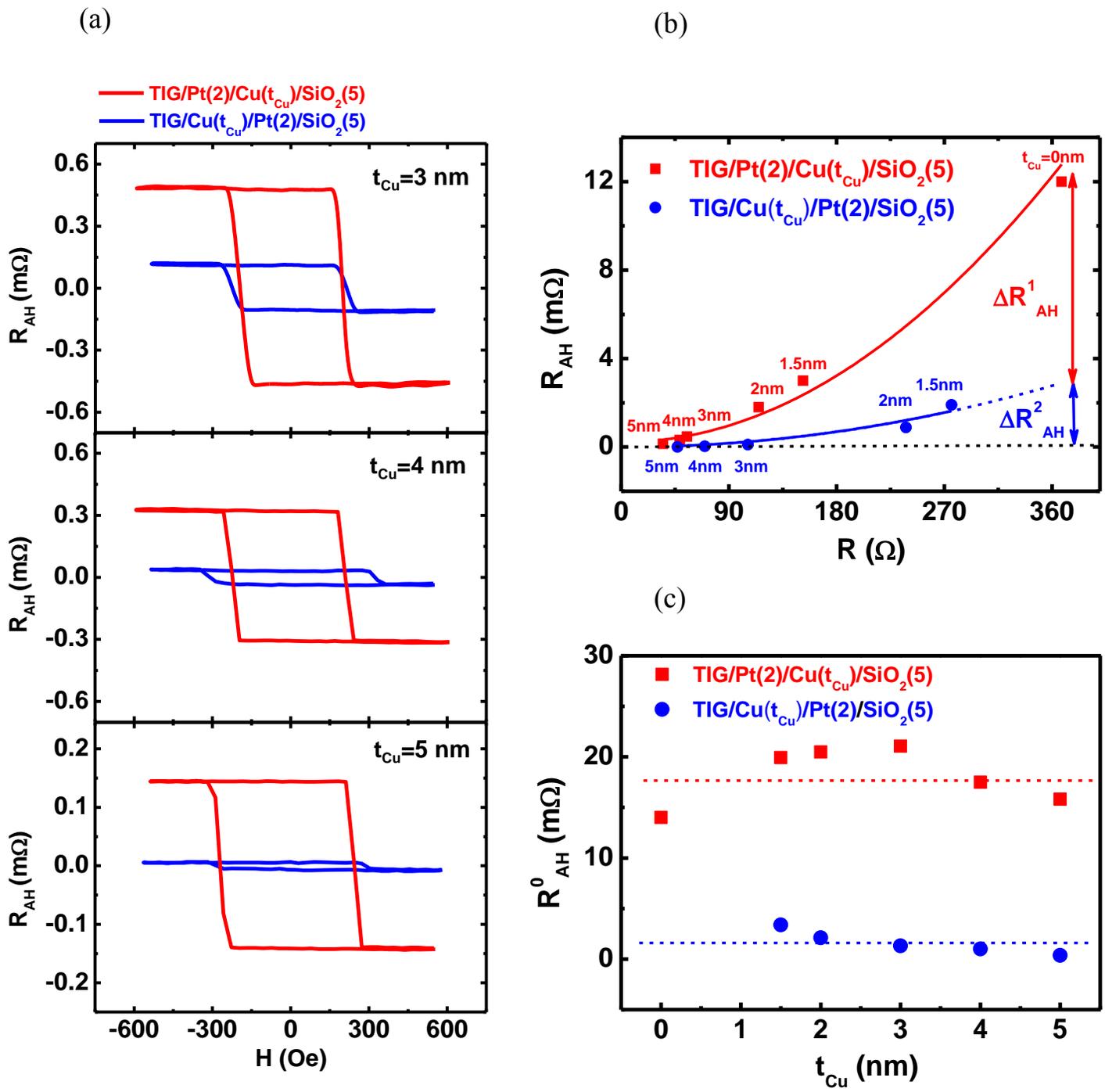

**Figure 4**